\documentstyle[aps,prl,twocolumn]{revtex}
\tighten
\begin{document}
\draft
\title
{Perturbative Inaccessibility of Conformal Fixed Points in Nonsupersymmetric
Quiver Theories}
\author
{Paul H. Frampton $^{(1,2)}$, and Peter Minkowski $^{(1,3)}$}
\address{ $^{(1)}$ TH Division, CERN, CH1211 Geneva 23, Switzerland, and}
\address
{$^{(2)}$ University of North Carolina,Chapel Hill, NC 27599, USA. }
\address
{$^{(3)}$ Institute for Theoretical Physics, University of Bern, Bern, Switzerland.}
\maketitle
\abstract
{The possibility that non-supersymmetric quiver theories may have
a renormalization-group fixed point at which there is conformal
invariance requires non-perturbative information.}

\medskip
\bigskip
\medskip

The AdS/CFT correspondence of Maladacena\cite{Mald} 
provides a powerful tool to study 
non-gravitational gauge field theories in flat {\it e.g.}
four-dimensional spacetime. In particular, it
has been suggested \cite{F} that nonsupersymmetric
quiver gauge theories obtained from compactification
of the Type IIB superstring on $AdS_5 \times S^5/\Gamma$
may, for finite number $N$ of $D3-$branes, possess
a conformal fixed point at the TeV scale and that
one such theory may be the correct path to go beyond
the standard model of particle phenomenology.

Three years ago, in a paper\cite{CST}
an argument based on perturbation
theory was used 
to criticize this whole approach to string phenomenology.
Our purpose here is not just to comment on that paper but
on what we perceive to be a widespread belief that the way
to approach the issue is through perturbative analysis.

Here we shall show that the S-duality of the underlying 
IIB superstring constrains the relevant
RG beta function $\beta(\alpha)$ to
suggest a conformal fixed point but that
to any order of perturbation theory such a fixed point
must remain inaccessible. We shall
show this by an illustration of the
possible form of $\beta(\alpha)$
which may be analytic around $\alpha = 0$
and $\alpha = \infty$ as a function of $\alpha$
with arbitrary coefficients in the Taylor
expansion for small $\alpha$.

Let us assume that the quiver theory has one
independent coupling constant which is asymptotically free
for small $\alpha$ so that the perturbative expansion is
(here $t = {\rm ln} \mu$)
\begin{equation}
\beta = \frac{d\alpha}{dt} = b_0 \alpha^2 + b_1 \alpha^3 + b_2 \alpha^4 + ...
\label{pert}
\end{equation}
with $b_0 < 0$. As already mentioned in \cite{FV} the underlying S-duality
under $\alpha \rightarrow 1/\alpha$ implies by
\begin{equation}
\beta(1/\alpha) = - \frac{1}{\alpha^2} \beta(\alpha)
\label{SDUAL}
\end{equation}
that $\beta(1) = 0$.

Let us illustrate the solution of Eq.(\ref{SDUAL})
in the form
\begin{equation}
\beta(\alpha) = - \alpha (\alpha - \frac{1}{\alpha}) F(\alpha)
\end{equation}
where $F(\alpha) = + F(1/\alpha)$. In particular consider
\begin{equation}
F(\alpha) = 
\Sigma_{n=0}^{n=\infty} C_n \alpha^{n+2} (1 + \alpha^{2n+4})^{-1}
\end{equation}
Then
\begin{equation}
\beta(\alpha) = (1 - \alpha^2)
[ C_0 \alpha^2 (1 + \alpha^4)^{-1}
+ C_1 \alpha^3 (1 + \alpha^6)^{-1} + ...]
\end{equation}
so that
\begin{equation}
\beta(\alpha) = C_0 \alpha^2 + C_1 \alpha^3 + (C_2 - C_0) \alpha^4 + ....
\end{equation}
can reproduce any expansion of
the perturbative type Eq.(\ref{pert}) by
appropriate choice of the coefficients $C_n$.
This form for $\beta(\alpha)$ can be analytic
for both $|\alpha| < 1$ and $|\alpha| > 1$ 
but must be singular at least somewhere 
on the unit circle $|\alpha| = 1$.
In particular, 
$\beta(\alpha)$ can be
analytic and vanishing at the points $\alpha = 0 ~ {\rm and}
~ \infty$.
The specific illustration chosen has a simple zero 
on the real axis at $\beta(1) = 0$.
It is clear that analysis based on perturbative calculation
of the coefficients $b_n$ in Eq.(\ref{pert}) could neither prove nor disprove
the existence of this zero in $\beta$.

The point is that while calculation of the perturbative expansion
for $\beta(\alpha)$ in Eq.(\ref{pert}) may well lead to supporting
evidence (but not rigorous proof) of a fixed point for weak coupling, as in {\it e.g.}
the well-known example\cite{BZ} of QCD with 16 flavors, such a calculation 
can never by itself confirm the 
presence or absence of a conformal fixed point.

\bigskip
\bigskip

The work of PHF was supported in part by the
Office of High Energy, US Department
of Energy under Grant No. DE-FG02-97ER41036.
We thank Luis Alvarez-Gaume for a discussion.

\end{document}